\documentclass[aps,prd,twocolumn,amsmath,showpacs,superscriptaddress,floatfix,nofootinbib,
nopreprintnumbers]{revtex4-1}

\usepackage{verbatim}
\usepackage[T1]{fontenc}
\usepackage[utf8]{inputenc}
\usepackage[american]{babel}
\usepackage{epsfig}
\usepackage{graphicx}
\usepackage{booktabs}
\usepackage{multirow}
\usepackage{dcolumn}
\usepackage{amsmath}
\usepackage{mathtools}
\usepackage{amsfonts}
\usepackage{amssymb}
\usepackage{epstopdf}
\usepackage{siunitx}
\usepackage{braket}
\usepackage{enumitem}
\usepackage{soul}
\usepackage[table]{xcolor}
\usepackage{color}
\usepackage{transparent}
\usepackage{pifont}
\usepackage[normalem]{ulem}

\usepackage{hyperref}
\hypersetup{
    colorlinks=true, 
    linkcolor=blue, 
    citecolor=magenta}

\newcolumntype{R}{>{\raggedleft\arraybackslash}p{0.6cm}} 

\newcommand\ev{\mathrm{eV}}

\newcommand{\diff}{\mathop{}\!\mathrm{d}}

\usepackage{booktabs}
\usepackage{multirow}
\usepackage{dcolumn}
\usepackage{colortbl}

\definecolor{magenta(process)}{rgb}{1.0, 0.0, 0.56}

\definecolor{darkspringgreen}{rgb}{0.09, 0.45, 0.27}

\DeclarePairedDelimiter{\abs}{\lvert}{\rvert}

\begin{document}

\title{How much do neutrinos live and weigh?} 

\author{Federica Pompa}
\email{federica.pompa@ific.uv.es}
\affiliation{Instituto de F\'{i}sica Corpuscular (IFIC), University of Valencia-CSIC, Parc Cient\'{i}fic UV, c/ Cate\-dr\'{a}tico Jos\'{e} Beltr\'{a}n 2, E-46980 Paterna, Spain\\}

\author{Olga Mena}
\email{omena@ific.uv.es}
\affiliation{Instituto de F\'{i}sica Corpuscular (IFIC), University of Valencia-CSIC, Parc Cient\'{i}fic UV, c/ Cate\-dr\'{a}tico Jos\'{e} Beltr\'{a}n 2, E-46980 Paterna, Spain\\}

\date{\today}
%
%
\begin{abstract}
The next-generation water Cherenkov Hyper-Kamiokande detector will be able to detect thousands of neutrino events from a galactic Supernova explosion via Inverse Beta Decay processes followed by neutron capture on Gadolinium. 
This superb statistics provides a unique window to set bounds on neutrino properties, as its mass and lifetime. 
We shall explore the capabilities of such a future detector, constraining the former two properties via the time delay and the flux suppression induced in the Supernovae neutrino time and energy spectra. 
Special attention will be devoted to the statistically sub-dominant elastic scattering induced events, normally neglected, which can substantially improve the neutrino mass bound via time delays. 
When allowing for a invisible decaying scenario, the $95\%$~C.L. lower bound on $\tau/m$ is almost one order of magnitude better than the one found with SN1987A neutrino events. 
Simultaneous limits can be set on both $m_\nu$ and $\tau_{\nu}$, combining the neutrino flux suppression with the time-delay signature: the best constrained lifetime is that of $\nu_1$, which has the richest electronic component. We find $\tau_{\nu_1}\gtrsim 4\times 10^5$~s at $95\%$~C.L. 
The tightest $95\%$~C.L. bound on the neutrino mass we find is $0.34~\ev$, which is not only competitive with the tightest neutrino mass limits nowadays, but also  comparable to future laboratory direct mass searches. 
Both mass and lifetime limits are independent on the mass ordering, which makes our results very robust and relevant.
 
\end{abstract}
\maketitle
\section{Introduction}

The detection of core-collapse Supernova (SN) neutrinos from a future galactic explosion could provide measurements and/or bounds on an incredibly broad number of neutrino properties, such as their mass~\cite{Pagliaroli:2010ik,Loredo:2001rx,Pompa:2022cxc}, mixing parameters~\cite{Takahashi:2001dc, Dighe:2003jg,Scholberg:2017czd,Hajjar:2023xae,Brdar:2022vfr}, Earth  matter effects in the neutrino propagation~\cite{Lagage:1987xu,Arafune:1987cj,Notzold:1987vc,Minakata:1987fj,Smirnov:1993ku,Dighe:1999bi,Lunardini:2000sw,Takahashi:2000it,Lunardini:2001pb,Takahashi:2001dc,Fogli:2001pm,Lindner:2002wm,Lunardini:2003eh,Dighe:2003jg,Dighe:2003vm,Akhmedov:2005yt,Dasgupta:2008my,Guo:2008mma,Scholberg:2009jr,Borriello:2012zc,Hajjar:2023knk}, decays or non-standard interactions beyond the Standard Model (SM) paradigm~\cite{Shalgar:2019rqe,Delgado:2021vha,Suliga:2020jfa}, as well as properties of the progenitor star~\cite{Brdar:2018zds} or the Supernova neutrino models~\cite{Hyper-Kamiokande:2021frf}. 
In this regard, future neutrino detectors, generally devoted to study the neutrino mixing unknowns exploiting accelerator neutrinos, can also act as Supernova neutrino observatories.
The future Hyper-Kamiokande (HK)~\cite{Hyper-Kamiokande:2018ofw,Hyper-Kamiokande:2021frf} water Cherenkov detector will detect over ten thousand neutrino events, for SN explosions occurring in our galaxy within a few tens of kiloparsecs. 
Therefore, it is timely to explore the sensitivities of such a future facility to neutrino properties. 
Here we shall focus on the neutrino mass and lifetime. 
Concerning the mass, we exploit the time delay experienced by neutrinos after traveling a distance $D$ to the Earth~\cite{Zatsepin:1968kt}
\begin{equation}
    \label{eq:t_delay}
    \Delta t = \frac{D}{2c}\left(\frac{m_\nu}{E_{\nu}}\right)^2~,
\end{equation}
\noindent see Refs.~\cite{Pagliaroli:2010ik,Loredo:2001rx} for early work and the more recent analysis of Ref.~\cite{Pompa:2022cxc} performed with the future DUNE Liquid Argon detector, where a sub-$\ev$ bound on $m_\nu$ has been derived. 
This very competitive limit, comparable to those expected from laboratory direct neutrino mass searches, was based on the detection of the SN neutronization peak in the electron neutrino time spectrum. 
Thanks to the large statistics expected in HK and the possibility of combining statistics from different detection channels, the former bound will be improved, as we shall see in the following. 
Concerning neutrino decays, in case the daughter neutrino is a sterile state, the expected neutrino flux is suppressed by the energy-dependent factor~\cite{Beacom:2002vi}
\begin{equation}
f_i = \exp{\left( - \frac{D}{E }\frac{m_{\nu_i}}{\tau_{\nu_i}}\right)}~,
\label{eq:fi}
\end{equation}
\noindent where $m_{\nu_i}$ and $\tau_{\nu_i}$ refer to the mass and lifetime of the neutrino mass eigenstate $i$, respectively. 
While current bounds on neutrino lifetimes are always quoted in terms of $\tau_{\nu_i}/m_{\nu_i}$ due to the impossibility of extracting independently these two parameters, we shall see that it is possible to constrain both $m_{\nu_i}$ and $\tau_{\nu_i}$ separately, exploiting neutrinos from SN. 

The structure of the paper is as follows. Section~\ref{sec:fluxes} describes the SN neutrino fluxes and the expected events in HK for different detection channels and neutrino flavors.  In Sec.~\ref{sec:masssec} we exploit the time delay of SN neutrinos to set a bound on the neutrino mass from HK, neglecting the effects of a possible neutrino decay and exploring different detection channels. Section~\ref{sec:masstausec} presents the combined analysis for both masses and lifetimes. Finally, we draw our conclusions in Sec.~\ref{sec:conclusions}.

\section{Supernova neutrinos}
\label{sec:fluxes}
Core-collapse SNe are factories of $\mathcal{O}(\SI{10}{\mega\electronvolt})$ neutrinos of all flavors. 
Their explosion mechanism can be divided into three phases: the \emph{neutronization burst}, which lasts for about $\SI{25}{\milli\second}$ and is characterized by a huge emission of electron neutrinos ($e^- + p\rightarrow \nu_e + n$); the \emph{accretion phase}, lasting $\sim \SI{0.5}{\second}$ during which high luminosity $\nu_e$ and $\bar{\nu}_e$ fluxes are radiated via the processes $e^- + p\rightarrow \nu_e + n$ and $e^+ + n \rightarrow \bar{\nu}_e + p$, and the final \emph{cooling phase}, when a hot neutron star is formed following the emission of fluxes of neutrinos and anti-neutrinos of all species within $\mathcal{O}(\SI{10}{\second})$.
We use the following quasi-thermal parameterization, representing well detailed numerical simulations \cite{Keil:2002in,Mirizzi:2015eza,Hudepohl:2009tyy,Tamborra:2012ac} 
\begin{equation}
\label{eq:differential_flux}
\Phi^{0}_{\nu_\beta}(t,E) = \frac{L_{\nu_\beta}(t)}{4 \pi D^2}\frac{\varphi_{\nu_\beta}(t,E)}{\langle E_{\nu_\beta}(t)\rangle}\,,
\end{equation}
and describing the double differential flux for each neutrino flavor $\nu_\beta$ of energy $E$ at a time $t$ after the SN core bounce, located at a distance $D$. 
Here, $L_{\nu_\beta}(t)$ is the $\nu_\beta$ luminosity, $\langle E_{\nu_\beta}(t)\rangle$ the mean neutrino energy and $\varphi_{\nu_\beta}(t,E)$ is the normalized neutrino energy distribution, defined as:
\begin{equation}
\label{eq:nu_energy_distribution}
\varphi_{\nu_\beta}(t,E) = \xi_\beta(t) \left(\frac{E}{\langle E_{\nu_\beta}(t)\rangle}\right)^{\alpha_\beta(t)} \exp{\left\{\frac{-\left[\alpha_\beta(t) + 1\right] E}{\langle E_{\nu_\beta}(t)\rangle}\right\}},
\end{equation}
where $\alpha_\beta(t)$ is a \emph{pinching} parameter and $\xi_\beta(t)$ is a unit-area normalization factor.
Input values for luminosity, mean energy and pinching parameter have been taken from the \texttt{SNOwGLoBES} software \cite{snowglobes}. \texttt{SNOwGLoBES} includes fluxes from the Garching Core-Collapse Modeling Group~\footnote{\url{https://wwwmpa.mpa-garching.mpg.de/ccsnarchive/index.html}}, providing simulation results for a progenitor star of $8.8 M_\odot$~\cite{Hudepohl:2009tyy}.
Neutrinos undergo flavor conversion inside the SN via the MSW (Mikheyev-Smirnov-Wolfenstein)  matter effect~\cite{Dighe:1999bi}. The neutrino fluxes at the Earth surface ($\Phi_{e}$ and $\Phi_{\mu}=\Phi_{\tau}=\Phi_{x}$) read as: 
\begin{align}
\Phi_e &= p\, \Phi_e^0 + (1-p) \Phi_x^0 \,, \label{eq:phi_e_earth} \\
\Phi_x &= \frac{1}{2} \big[(1-p) \Phi_e^0 + (1 + p) \Phi_x^0\big] \,, \label{eq:phi_x_earth}
\end{align}
with $\Phi^0$ being the neutrino flux produced in the SN core and $p$ the oscillation probability, with $p = \abs{U_{e1}}^2$ for antineutrinos in Normal Ordering (NO), $p = \abs{U_{e2}}^2$ for neutrinos in Inverted Ordering (IO), and $p = \abs{U_{e3}}^2$ for neutrinos~(antineutrinos) in NO~(IO). 
Notice that here we neglect possible non-adiabaticity effects occurring when the resonances occur near the shock wave \cite{Schirato:2002tg,Fogli:2003dw,Fogli:2004ff,Tomas:2004gr,Dasgupta:2005wn,Choubey:2006aq,Kneller:2007kg,Friedland:2020ecy}, and the presence of turbulence in the matter density \cite{Fogli:2006xy,Friedland:2006ta,Kneller:2010sc,Lund:2013uta,Loreti:1995ae,Choubey:2007ga,Benatti:2004hn,Kneller:2013ska,Fogli:2006xy}. Also, effects due to interactions on Earth matter can be ignored \cite{Pompa:2022cxc}. 

The main interaction channel in the HK detector is Inverse Beta Decay (IBD): 
\begin{equation}
\label{eq:IBD}
\bar{\nu}_{e} + p \longrightarrow n + e^{+}\,.
\end{equation}
While the emitted positron is promptly detected by HK, the neutron must thermalize and be captured on a proton before being detected. 
Following the capture (after $\sim\SI{200}{\micro\second}$), a gamma-ray of $\SI{2.2}{\mega\electronvolt}$ is emitted, but it cannot be detected because of the HK efficiency cut.
However, by adding Gadolinium to the water, the emitted neutron is captured in $\sim\SI{20}{\micro\second}$, producing a $\gamma$-ray signal with an energy of $\sim\SI{8}{\mega\electronvolt}$,  that can be observed by HK.
This allows tagging, with an efficiency near $90\%$, IBD events and distinguish them from those events coming from other interaction channels.

\begin{figure}
    \centering
\includegraphics[width=\columnwidth]{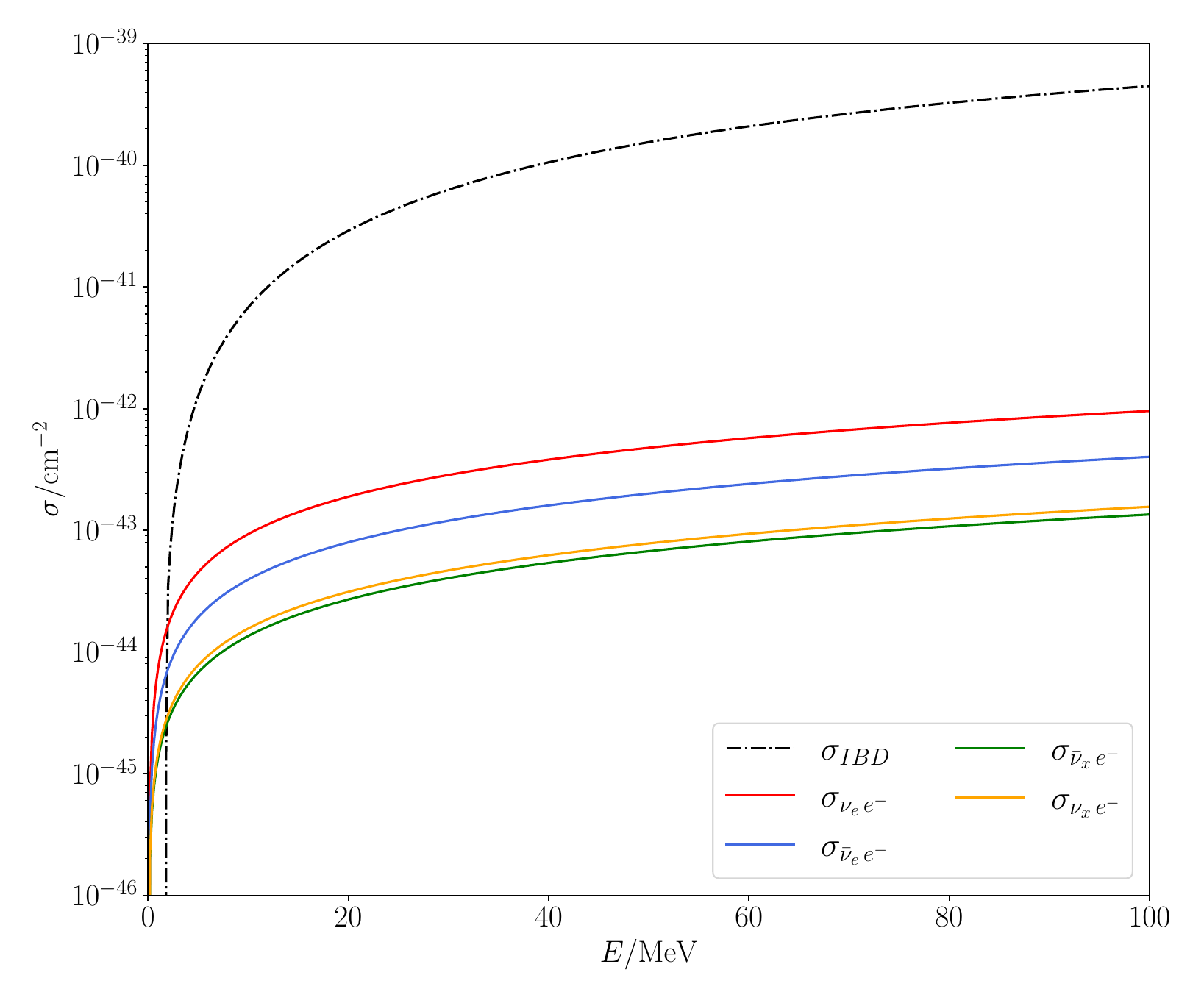}
    \caption{Cross-sections of the relevant processes for SN core-collapse neutrinos in HK: IBD and ES. The ES ones have been taken from \texttt{SNOwGLoBES} \cite{snowglobes}.}
    \label{fig:IBD_and_ES_xsecs}
\end{figure}

\begin{figure}
    \centering
\includegraphics[width=\columnwidth]{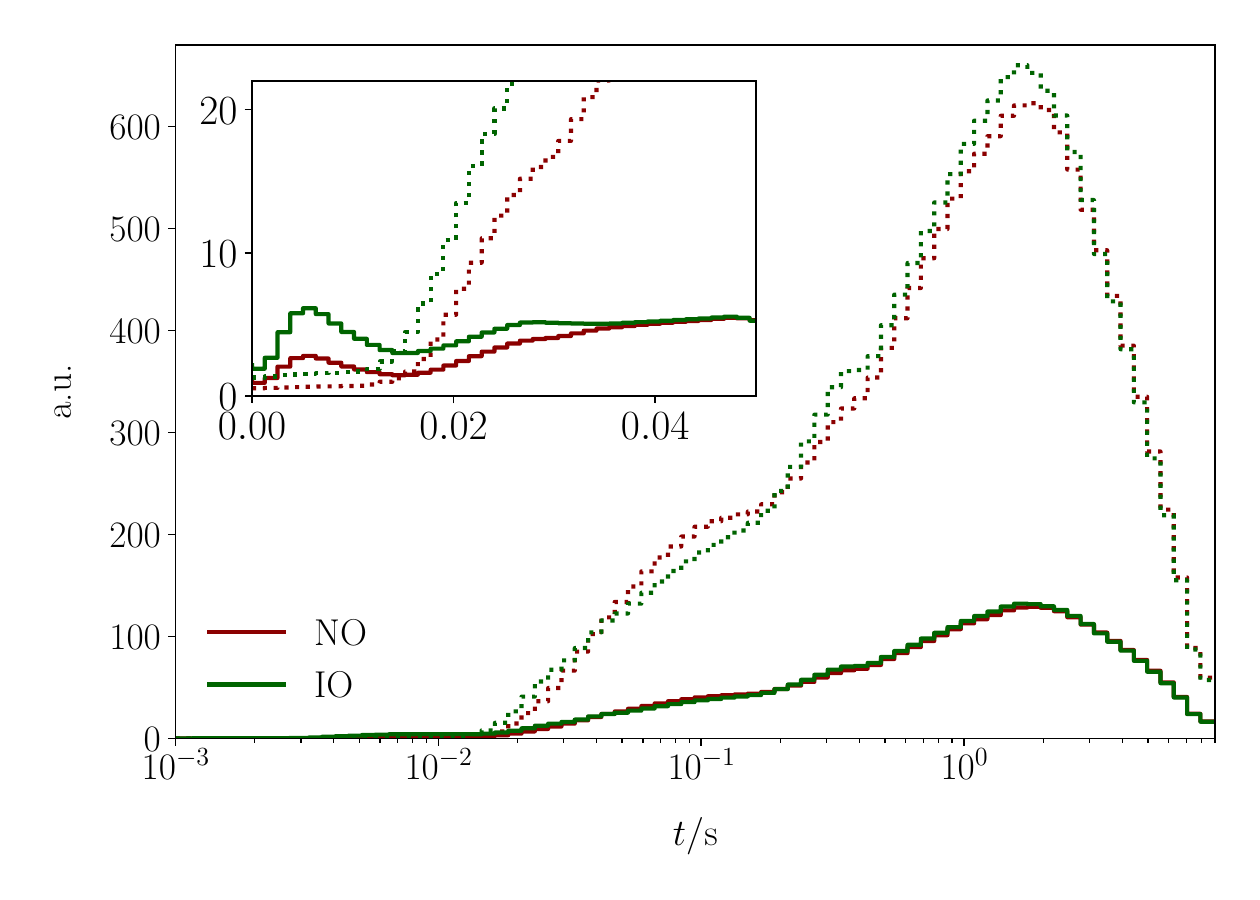}
    \caption{Number of neutrino events coming from the $90\%$ of Gd-tagged IBD (dotted lines) and those from the ES plus untagged IBD (solid lines) interactions in HK as a function of time.  
    The inset depicts the number of neutrino events expected in HK during the first $\SI{50}{\milli\second}$ of the SN burst.}
    \label{fig:rate_in_time}
\end{figure}

\begin{figure}
    \centering
    \includegraphics[width=\columnwidth]{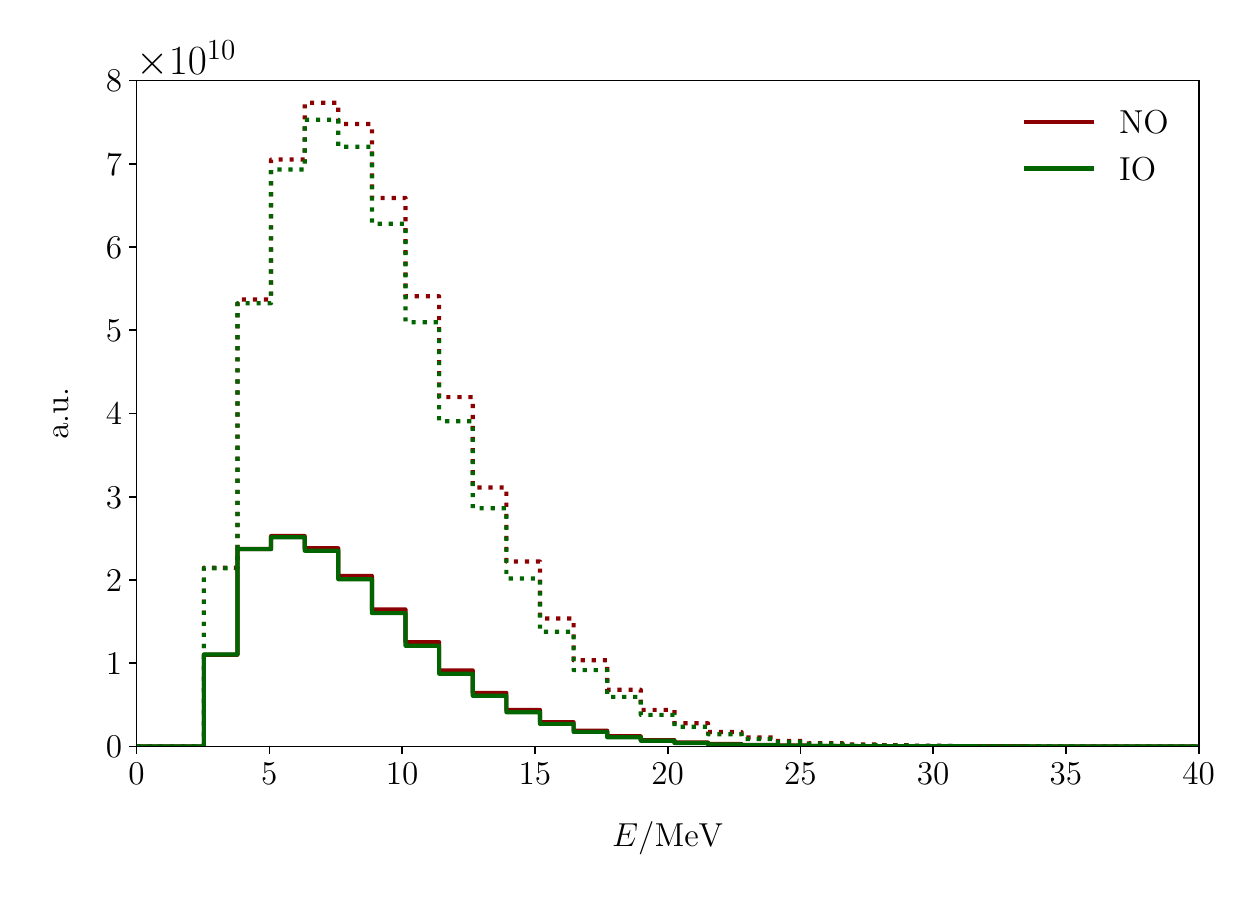}
    \caption{Number of neutrino events coming from the $90\%$ of Gd-tagged IBD (dotted lines) and those from the ES plus untagged IBD (solid lines) interactions in HK as a function of energy.}
    \label{fig:rate_in_energy}
\end{figure}

Elastic scatterings (ES) of neutrinos of all flavors with electrons: 
\begin{equation}
\label{eq:ES}
\nu + e \longrightarrow \nu + e\,,
\end{equation}
constitute a subdominant channel in terms of statistics. 
The scattered electrons are forward-peaked, so an angular cut can be applied in order to detect the majority of them. 
The different cross-sections, versus the SN neutrino energy, for the interaction channels here mentioned, have been taken from \texttt{SNOwGLoBES} \cite{snowglobes} and are depicted in Fig.~\ref{fig:IBD_and_ES_xsecs}. 
The expected number of events coming from a SN explosion at $10$~kpc from Earth (which will be the fiducial SN location throughout this manuscript) for each interaction channel, (anti)neutrino flavor and oscillation scenario, is shown in Tab.~\ref{tab:rates_HK}. Notice that the addition of Gd allows tagging the $90\%$ of the total IBD events. The $10\%$ IBD events remaining cannot be distinguished from other interactions and must be included in the ES count.~\footnote{The $10\%$ inefficiency to tag IBD events is taken to be independent of neutrino energy and emission time.} No angular cut has been applied to the ES channel.
Figures~\ref{fig:rate_in_time} and \ref{fig:rate_in_energy} show all these numbers of events as a function of the emission time and neutrino energy, respectively. 

\begin{table}
    \centering
    \caption{Total number of predicted events in each interaction channel in the two neutrino mass orderings. The relevant ones for the analysis are obtained following the addition of Gd, which allows tagging the $90\%$ of total IBD, while the remaining $10\%$ cannot be distinguished from other interactions and must be included in the ES count.}
    \label{tab:rates_HK}
    \begin{tabular}{cccc}
    \toprule
    Channel & flavor & NO & IO \\
    \midrule
    Gd-tagged IBD  & $\bar{\nu}_e$  & $16209$ & $16666$ \\
    \midrule
    IBD untagged  & $\bar{\nu}_e$  & $1801$ & $1852$ \\
    \midrule
    \multirow{4}*{ES}   & $\nu_e$  & $786$ & $814$ \\
                        & $\bar{\nu}_e$ &  $328$ & $331$ \\	
                        & $\nu_x$ & $139$ & $137$ \\	
                        & $\bar{\nu}_x$ & $113$ & $112$ \\	
    \midrule
    IBD untagged + total ES & all  & $3419$ & $3494$ \\ 
    \bottomrule 
    \end{tabular}
\end{table} 
\section{Neutrino mass bound}
\label{sec:masssec}
We start by describing the procedure to compute the HK sensitivities to the neutrino mass via Supernova neutrino time delays, assuming a degenerate neutrino mass spectrum. 
We first generate  our \textit{experimental} datasets for each channel and oscillation scenario, assuming massless neutrinos and perfect time resolution for our studies. 
On the other hand, we assume a given energy resolution in the (MeV) energy range of interest, and smear the neutrino energy of each generated event. 
The two free parameters constrained in our fit are the neutrino mass $m_\nu$ and an offset time $t_\text{off}$ \cite{Pagliaroli:2008ur} between the moment when the earliest SN neutrino reaches the Earth and the detection of the first event $i=1$. 
The fitted emission time $t_{i, \rm fit}$ for each event $i$ depend on these two fit parameters as follows:
\begin{equation}
\label{eq:emission_t}
t_{i, \rm fit} = \delta t_i  - \Delta t_{i}(m_\nu) + t_\text{off}\,,
\end{equation}
where $\delta t_i $ is the time at which the neutrino interaction $i$ is measured in HK (with the convention that $\delta t_1\equiv 0$ for the first detected event), $\Delta t_i(m_\nu)$ is the delay induced by the non-zero neutrino mass (see Eq.~\ref{eq:t_delay}), and $t_\text{off}$ is the offset time. 
This parameter depends on the neutrino interaction cross-section, so it is different for each detection channel; furthermore, the addition of Gd in water allows discriminating the IBD interactions from the rest, making the two channels (i.e. Gd-tagged IBD and untagged IBD + ES) statistically independent. 
This translates into the possibility of performing a separate minimization for each of them and then combine the results by summing. 
All in all, our likelihood function $\mathcal{L}$ reads as~\cite{Pagliaroli:2008ur}:
\begin{equation}
\label{eq:likelihood_fun}
\mathcal{L}(m_{\nu},t_\text{off}) = \prod_{i=1}^{N}\int R(t_i,E_i)G_i(E)\diff E~, 
\end{equation}
where $G_i$ is a Gaussian distribution with mean $E_i$ and  $\sigma= 0.6\sqrt{E_i}$, with $E_i$ in $\SI{}{\mega\electronvolt}$ units, accounting for the energy resolution and $R$ the rate function, defined as the convolution of the neutrino flux, the cross-section of the interaction, and on the detector efficiency.
In the analysis, the HK efficiency has been assumed like a step function with a energy threshold at $5~$MeV.
For each fixed value of $m_\nu$, we minimize the following $\chi^2$ function:
\begin{equation}
\label{eq:chi2_fun}
\chi^2(m_{\nu}) = -2 \log[\mathcal{L}(m_{\nu},t_\text{off,best})]~,
\end{equation}
where $\mathcal{L}(m_{\nu}, t_\text{off,best})$ indicates the maximum likelihood at this particular value of $m_\nu$. Finally, we combine 
all datasets for the same neutrino oscillation scenario to evaluate the impact of statistical fluctuations.  
For each value of $m_\nu$, we  compute the mean and the standard deviation of all toy dataset $\chi^2$ values. 
In order to estimate the allowed range in $m_\nu$, the $\Delta\chi^2$ difference between all mean $\chi^2$ values and the global mean $\chi^2$ minimum is computed.  The mean 95\% C.L. sensitivity to $m_\nu$ is then defined as the largest value of $m_\nu$ satisfying $\Delta \chi^2<3.84$. The $\pm 1\sigma$ uncertainty on the 95\% C.L. sensitivity to $m_\nu$ can be computed similarly, including into the $\Delta\chi^2$ evaluation also the contribution from the standard deviation of all toy dataset $\chi^2$ values.

\begin{figure}
\begin{tabular}{c}
 \includegraphics[width = 0.45\textwidth]
 {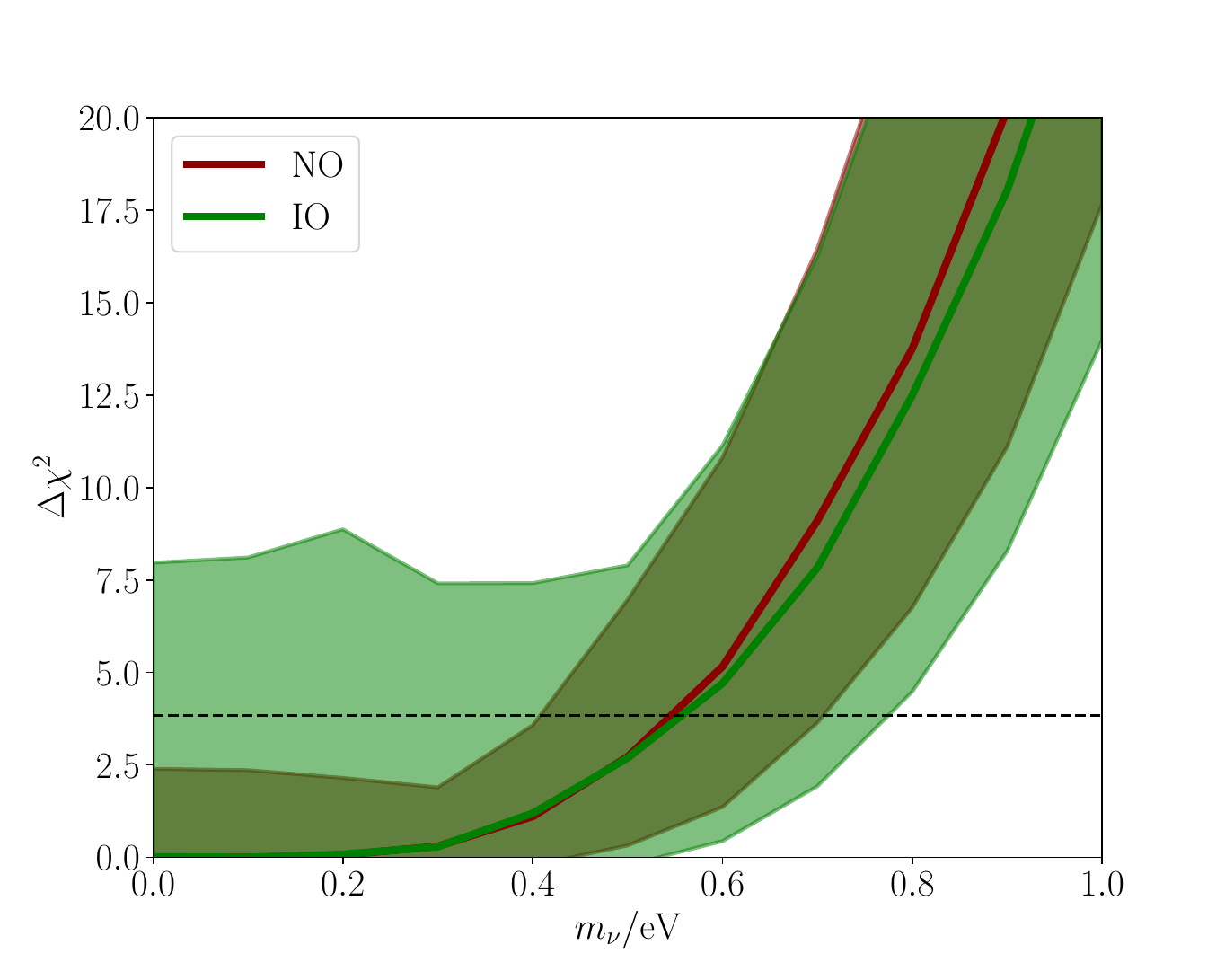}\\
 \includegraphics[width = 0.45\textwidth]
{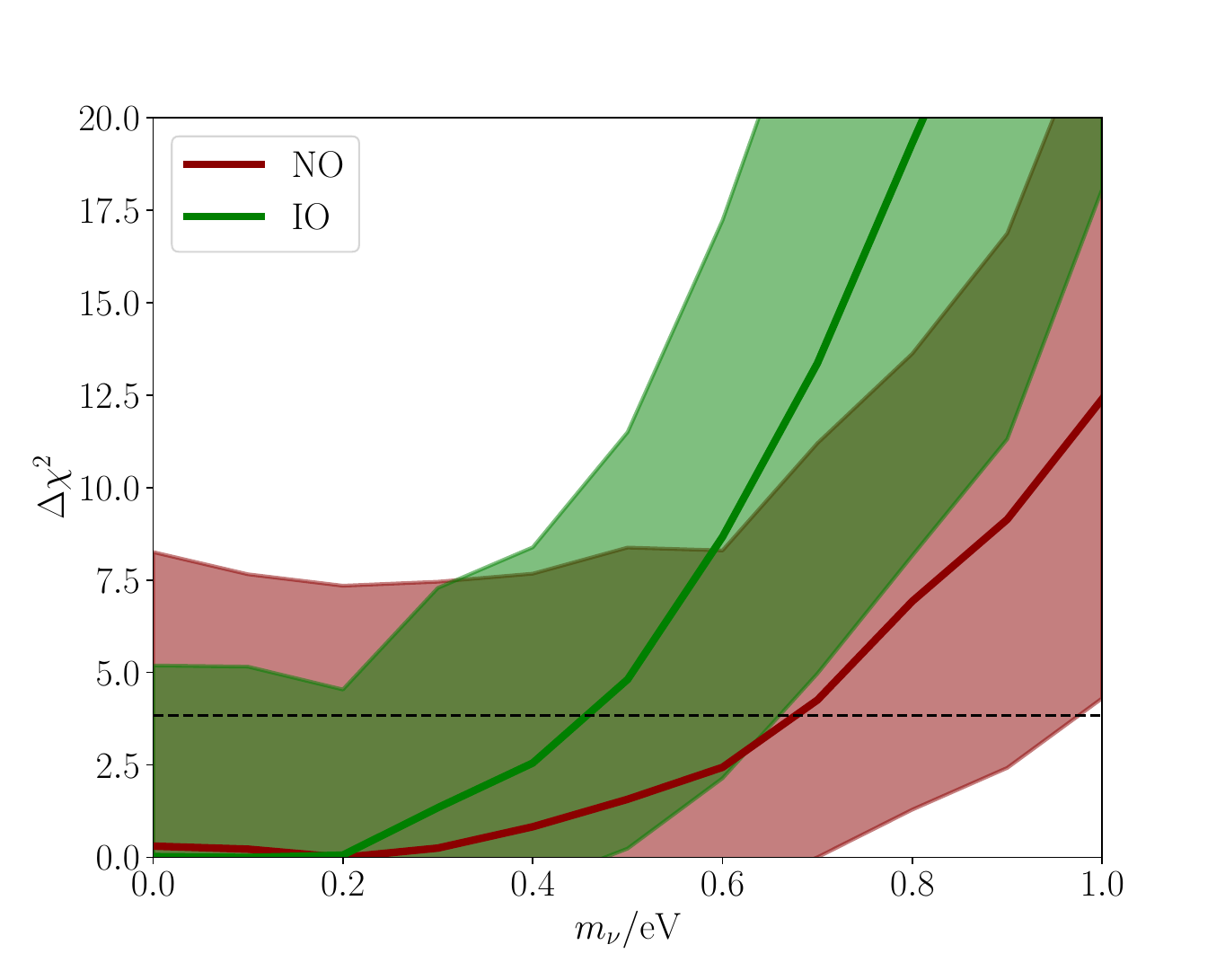}\\
 \includegraphics[width = 0.45\textwidth]
{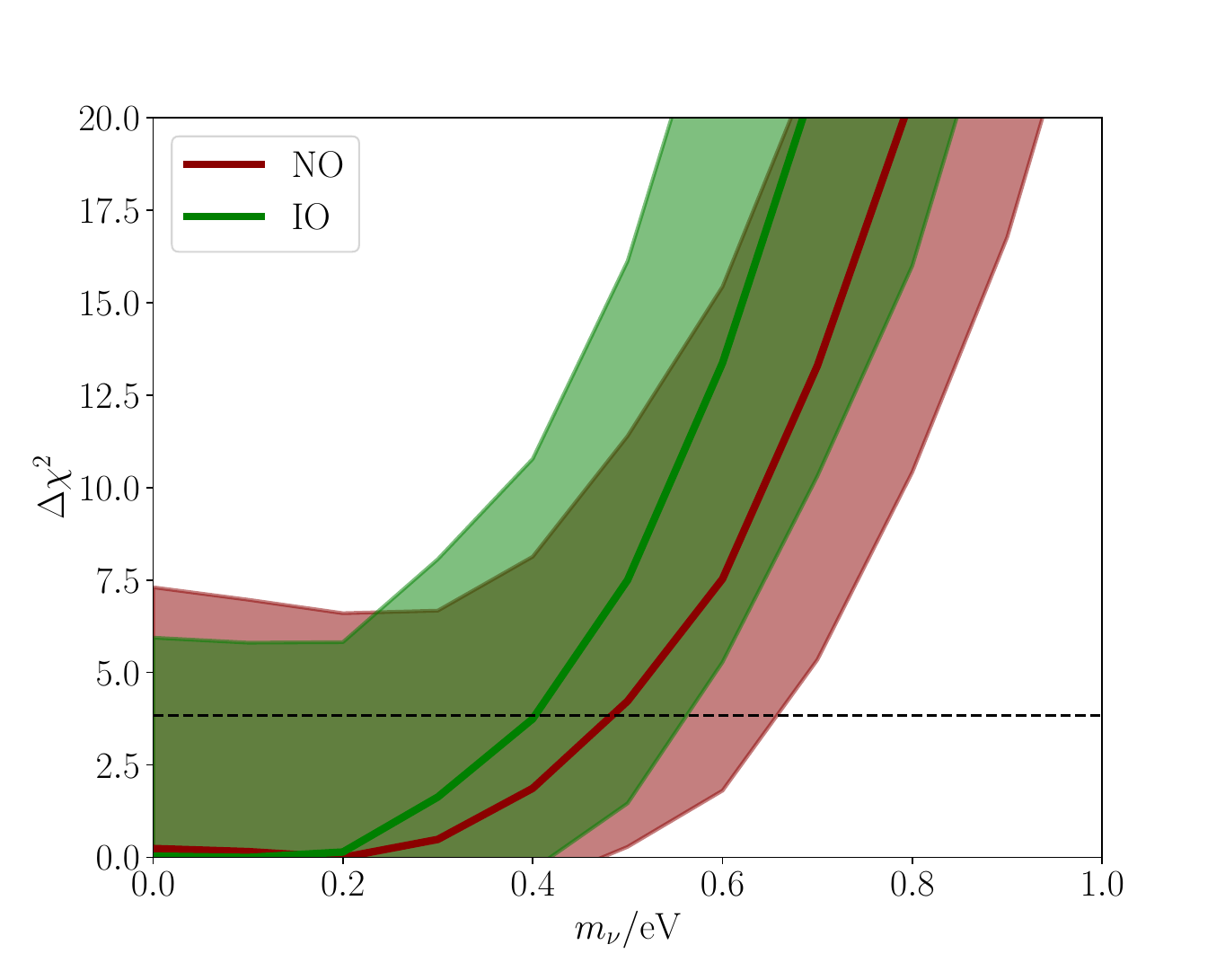}\\
\end{tabular}
    \caption{From top to bottom panel: $\Delta \chi^2$ as a function of $m_\nu$ for the likelihood exploited in the time-delay neutrino signal from $90\%$ of the total expected IBD events, from $10\%$ of the total expected IBD events (not tagged by Gd) plus the contribution of all flavors ES and from both the tagged IBD and the ES events, see the main text for details. The horizontal lines depict the $95\%$~C.L. sensitivities.}
\label{fig:mass}
\end{figure}

We show in Tab.~\ref{tab:mnu_bounds_HK} the $95\%$~C.L. upper bounds on the neutrino mass achieved in HK analyses in both individual and combined channels for the two possible oscillation schemes.  Figure~\ref{fig:mass} shows instead the $\Delta \chi^2$ profiles as a function of neutrino mass $m_\nu$, depicting the mean sensitivities and their $\pm 1\sigma$ uncertainties. Notice that,  by considering the Gd-tagged IBD case only (upper panel in Fig.~\ref{fig:mass}), due to the very high and similar statistics achieved in all the oscillation schemes (see Tab.~\ref{tab:rates_HK}), the expectation is to find a bound which is independent on the oscillation scenario. Indeed, this is what is obtained, see the first row of Tab.~\ref{tab:mnu_bounds_HK}. 

By adding the effect due to ES, which also allows being sensitive to the electron neutrino events from the SN neutronization burst (see the inset in Fig.~\ref{fig:rate_in_time}), the effects on neutrino mass bounds result to be relevant, moving to an oscillation-dependent global result (medium panel in Fig.~\ref{fig:mass}). This implies that, even if the expected number of events from the ES channel is smaller than the one coming from IBD, few events coming from the SN neutronization peak can set strong bounds on $m_\nu$~\cite{Pompa:2022cxc}. This is the reason why the best constraints are obtained in the oscillation scenarios in which this neutronization peak either remains unmodified or it is only partially suppressed, as it is the case for the IO scheme. To conclude the discussion, even if the ES channel is subdominant, it grants the observation of all flavors neutrinos, not only the $\bar{\nu}_e$, providing the possibility to observe also the SN neutronization peak and therefore improving considerably the limits on the neutrino mass based on time delay effects. 

Adding the two detection channels therefore results in an oscillation-dependent sensitivity limit on $m_\nu$: our tightest limits, which are obviously obtained within this combined channel scenario, are below $0.5$~eV, a factor of two better than those obtained in  Ref.~\cite{Pompa:2022cxc}, and quite similar for both mass orderings.
\begin{table}
    \centering
    \caption{$95\%$~C.L. sensitivity on $m_\nu$ in $\ev$, from a sample of HK SN datasets at $D=10$~kpc, for the two neutrino mass orderings.}
    \label{tab:mnu_bounds_HK}
    \begin{tabular}{ccccc}
    \toprule
    Channel &  & NO & & IO \\
    \midrule
    Gd-tagged IBD   &  & $0.54$ & & $0.56$ \\
    \midrule
    untagged IBD + total ES  & & $0.68$& & $0.46$ \\ 
    \midrule
    Combined  & & $0.48$ & & $0.40$ \\
    \bottomrule 
    \end{tabular}
\end{table} 
\section{Neutrino lifetime and mass joint constraints}
\label{sec:masstausec}
Since neutrinos are massive, they can decay. 
Radiative decays among two different neutrino mass eigenstates $\nu_j \rightarrow \nu_i +\gamma$~\cite{Petcov:1976ff,Marciano:1977wx} have lifetimes longer than the age of the universe, rendering these processes unobservable. 
Nevertheless, in the presence of a new light or massless mediator, such a decay rate could be higher. The equation describing the transition probabilities accounting also for invisible decays reads as~\cite{Beacom:2002vi}
 \begin{equation}
P_{\alpha \beta}=
\sum_{i=1}^3
|(U_{\rm PMNS})_{\alpha i}|^2
|(U_{\rm PMNS})_{\beta i}|^2
\exp{\left( - \frac{D}{E}\frac {m_{\nu_i}}{ \tau_{\nu_i}}\right)}~,
\label{eq:nu_decay}
\end{equation}
where $U_{\rm PMNS}$ is the three-neutrino mixing matrix, $\tau_{\nu_i}$ is the lifetime of the neutrino mass eigenstate $\nu_i$ in the rest frame, boosted in the laboratory frame by the factor $\gamma=E/m_{\nu_i}$.  Present lifetime limits on invisible neutrino decays range from CMB limits, $\tau_{\nu_i}/m_{\nu_i}> (2.6 - 0.6) \times 10^{10} (m_\nu /0.05 \ \textrm{eV})^2$~s/eV  at $95\%$~C.L.~\cite{Hannestad:2005ex,Archidiacono:2013dua,Escudero:2019gfk,Chacko:2020hmh,Escudero:2020ped}~\footnote{See however Refs.~\cite{Barenboim:2020vrr,Chen:2022idm} for possible scenarios where the former constraints are relaxed.}, to those found from atmospheric and long-baseline neutrino data, $\tau_{\nu_3}/m_{\nu_3} > 2.9 \times 10^{-10}$~s/eV at $90\%$~C.L.~\cite{Gonzalez-Garcia:2008mgl} (see Refs.~\cite{deSalas:2018kri,Choubey:2017dyu,Choubey:2020dhw} for future perspectives). 
Between these two limits we have, from less to more constraining bounds, those from solar neutrinos ($\tau_{\nu_2}/m_{\nu_2} > 2.27 \times 10^{-4}$~s/eV at $99\%$~C.L.~\cite{Joshipura:2002fb,Bandyopadhyay:2002qg}), from BBN ($\tau_{\nu_i}/m_{\nu_i} > 3 \times 10^{-3}$~s/eV at $95\%$~C.L.~\cite{Escudero:2019gfk}), from high-energy astrophysical neutrinos observed at IceCube ($\tau_{\nu_i}/m_{\nu_i}>10$~s/eV at $\gtrsim 2 \sigma$~\cite{Rasmussen:2017ert,Denton:2018aml,Abdullahi:2020rge,Baerwald:2012kc,Bustamante:2016ciw}) and from SN1987A neutrinos~\cite{Kamiokande-II:1989hkh,Kamiokande-II:1987idp,Bionta:1987qt,Alekseev:1988gp,Alekseev:1987ej}  on the electron antineutrino lifetime ($\tau_{\nu_i}/m_{\nu_i}>5.7 \times 10^5$~s/eV~\cite{Frieman:1987as}). 
Notice that, since the neutrino mass $m_{\nu_i}$ is unknown, lifetime limits are often quoted in terms of 
$\tau_{\nu_i}/m_{\nu_i}$. However, core-collapse SN neutrinos open the possibility of extracting simultaneously the mass and the lifetime, relying mostly on the time delays induced in the time spectrum and on the flux suppression induced in the energy spectrum, respectively. 

For the sake of simplicity, neglecting time delays, and assuming that all neutrino mass eigenstates are decaying, the overall picture of the neutrino flux arriving at the detector would be given, depending on the mass ordering, by
\begin{subequations}
    \begin{align}
        \Phi_e &= |U_{e3}|^2 f_3 \Phi_e^0 + [|U_{e1}|^2 f_1 + |U_{e2}|^2 f_2] \Phi_x^0 \label{eq:e_NO} \\
        \bar{\Phi}_e &= |U_{e1}|^2 f_1 \bar{\Phi}_e^0 + [|U_{e2}|^2 f_2 + |U_{e3}|^2 f_3] \Phi_x^0   \label{eq:ebar_NO} \\
        \Phi_x &= \frac{1}{2}[(1-|U_{e3}|^2 f_3) \Phi_e^0 + (2 - |U_{e1}|^2 f_1 - |U_{e2}|^2 f_2) \Phi_x^0)] \label{eq:x_NO} \\
        \bar{\Phi}_x &= \frac{1}{2}[(1 - |U_{e1}|^2 f_1) \bar{\Phi}_e^0 + (2 - |U_{e2}|^2 f_2 - |U_{e3}|^2 f_3) \Phi_x^0]   \label{eq:xbar_NO}
    \end{align}
    \label{eq:NO}
\end{subequations}
in NO, while
\begin{subequations}
    \begin{align}
        \Phi_e &= |U_{e2}|^2 f_2 \Phi_e^0 + [|U_{e1}|^2 f_1 + |U_{e3}|^2 f_3] \Phi_x^0 \label{eq:e_IO} \\
        \bar{\Phi}_e &= |U_{e3}|^2 f_3 \bar{\Phi}_e^0 + [|U_{e1}|^2 f_1 + |U_{e2}|^2 f_2] \Phi_x^0   \label{eq:ebar_IO} \\
        \Phi_x &= \frac{1}{2}[(1-|U_{e2}|^2 f_2) \Phi_e^0 + (2 - |U_{e1}|^2 f_1 - |U_{e3}|^2 f_3) \Phi_x^0] \label{eq:x_IO} \\
        \bar{\Phi}_x &= \frac{1}{2}[(1 - |U_{e3}|^2 f_3) \bar{\Phi}_e^0 + (2 - |U_{e1}|^2 f_1 - |U_{e2}|^2 f_2)) \Phi_x^0]   \label{eq:xbar_IO}
    \end{align}
    \label{eq:IO}
\end{subequations}
\noindent for IO, with $f_i$ given by Eq.~(\ref{eq:fi}). 
In the following, we shall explore both the time delay and the neutrino decay to see the combined effect and the sensitivities expected from the future HK detector. 
For that, one needs to convolute the fluxes given by the equations above with the time delay factor described in Eq.~(\ref{eq:t_delay}) and discussed in the previous section. To study the sensitivity to $\tau_\nu$ and $m_\nu$, the Poissonian $\chi^2$ function is adopted:
\begin{equation}
    \label{eq:poisson_chi2}
    \Delta \chi^2(m_\nu,\tau_\nu) = 2 \sum_i \Bigg[N_i - N_i^{\rm exp} + N_i^{\rm exp} \rm{ln} \frac{N_i^{\rm exp}}{N_i} \Bigg]
\end{equation}
with $N_i$ the rate content in the $i-$th energy and time bin assuming zero neutrino mass and no decay, while $N_i^{\rm exp}$ is the expected rate.~\footnote{For the sake of simplicity, we no longer consider the offset time $t_\text{off}$ as a parameter in the fit.}

\begin{figure*}
\begin{tabular}{c c c}
 \includegraphics[width = 0.33\textwidth]
{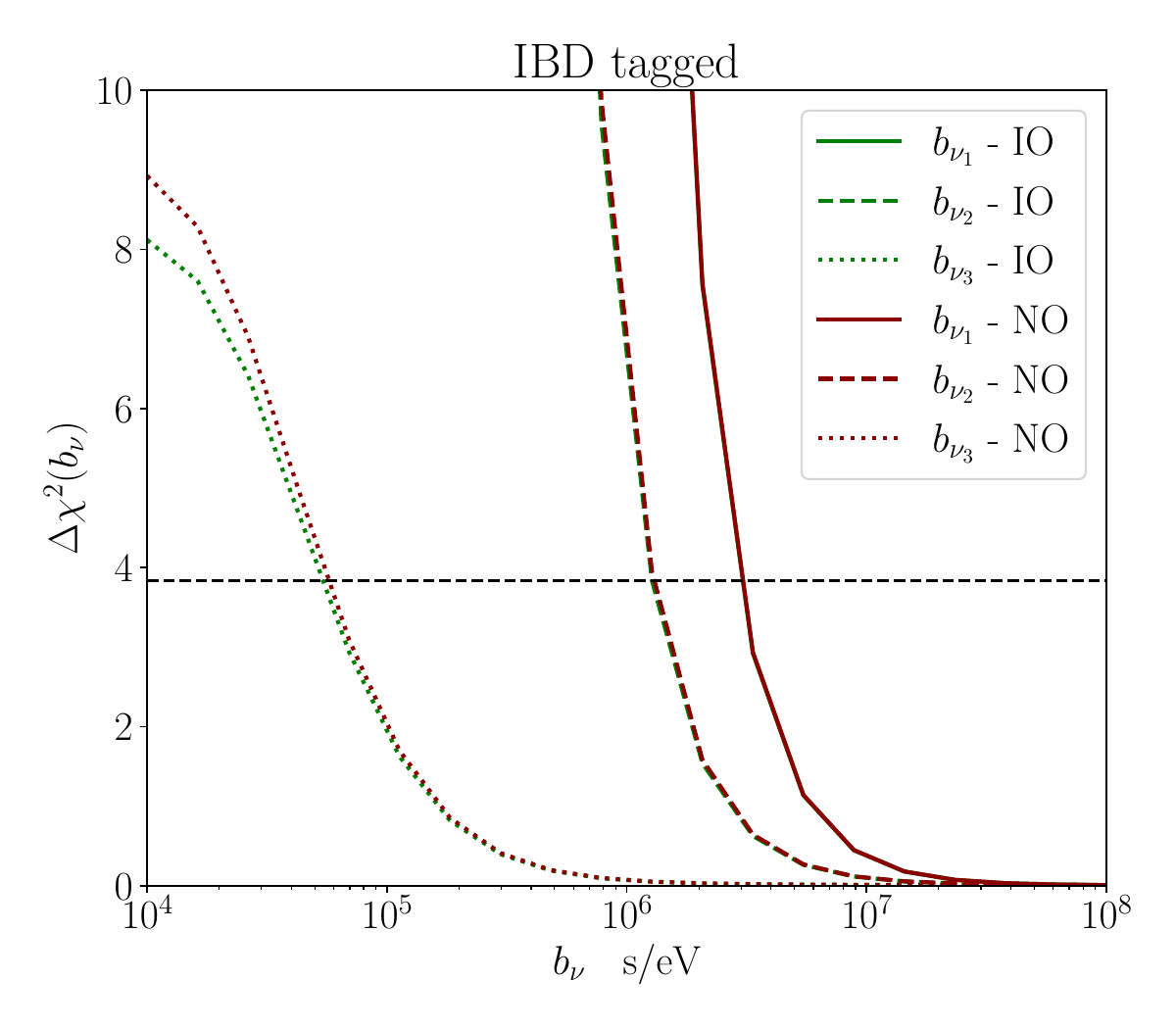}&
 \includegraphics[width = 0.33\textwidth]
{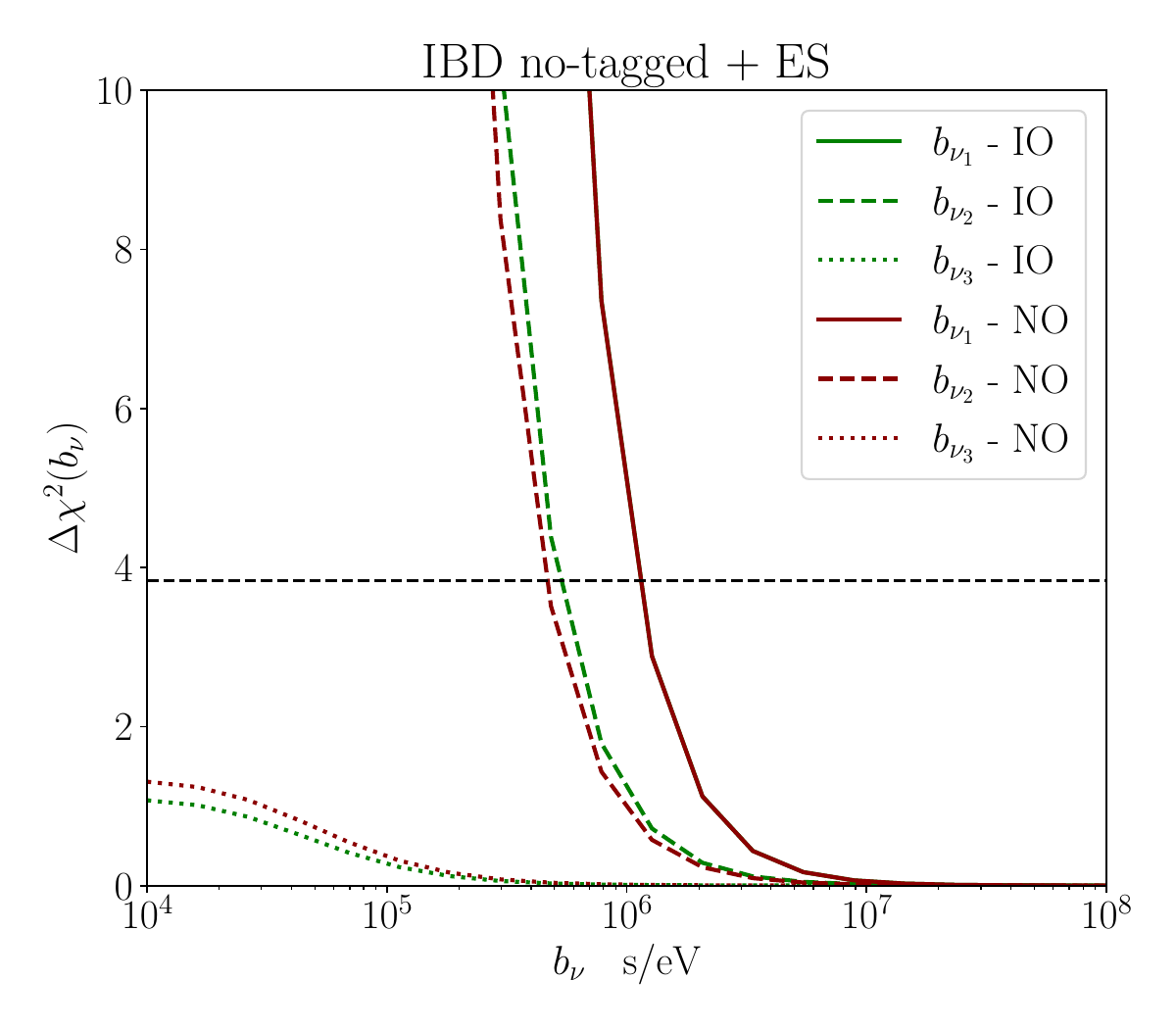}&
 \includegraphics[width = 0.33\textwidth]
{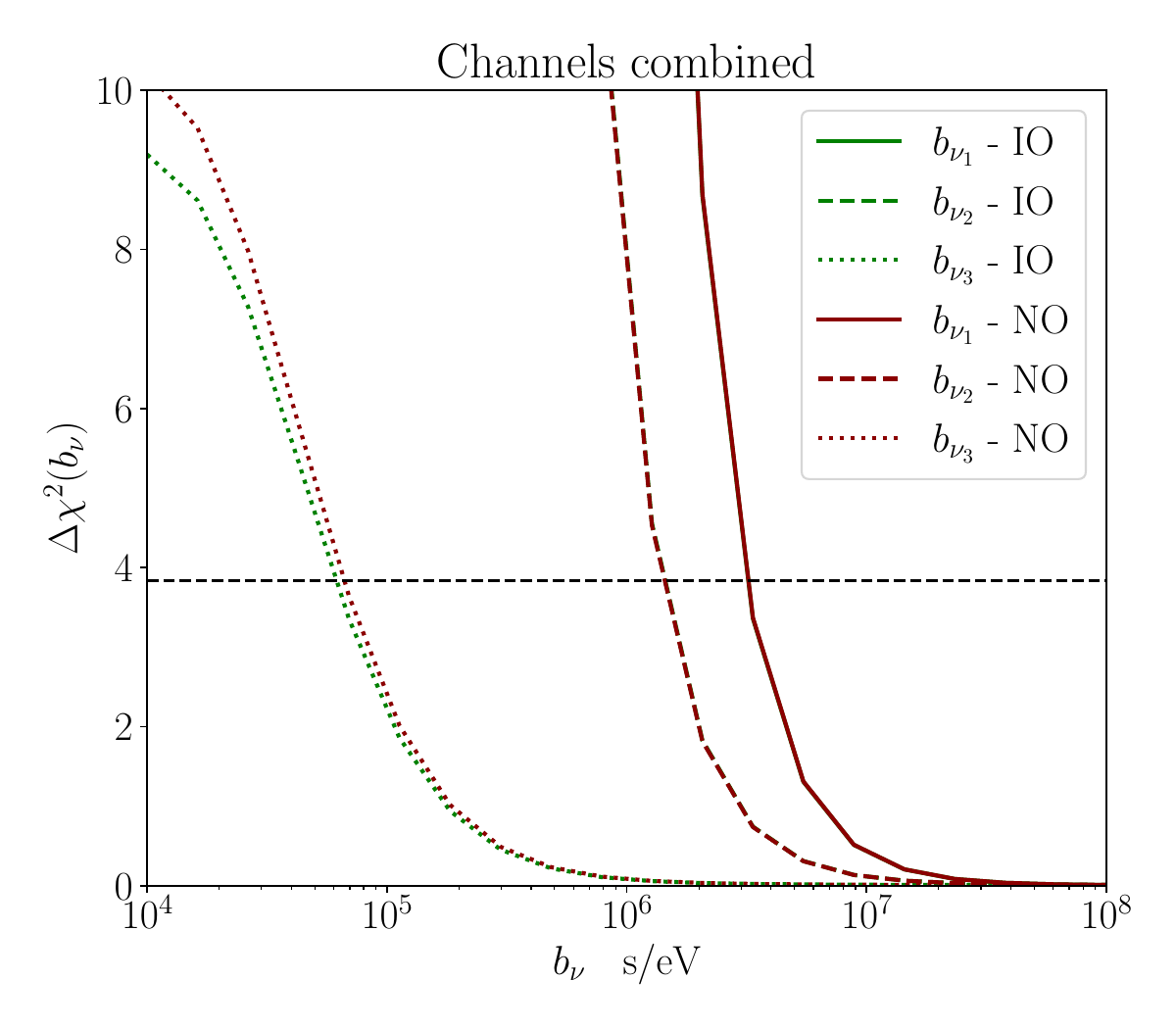}
\end{tabular}
\caption{From left to right panels: marginalized $\Delta \chi^2$ function versus $b_\nu = \frac{\tau_\nu}{m_\nu}$ for the likelihood exploited in the time-delay signal in decaying neutrino scenarios from $90\%$ of the total expected IBD events (tagged by Gd), from $10\%$ of the total expected IBD events (not tagged by Gd) plus the contribution of all-flavor ES, and from both the channels combined, see the main text for details. The horizontal lines depict the $95\%$~C.L. sensitivities.}
\label{fig:bparam}
\end{figure*}

We start by computing the limits on the quantity $b_\nu=\tau_\nu/m_\nu$, for the sake of comparison with existing limits. 
We assume in the following that the neutrinos are degenerate in mass. Table~\ref{tab:b_bounds_both_eff} and Fig.~\ref{fig:bparam} present the results for the different neutrino mass eigenstates arising from different detection channels. Notice that the limits barely depend on the hierarchy, and also that the tightest bounds are obtained for $\nu_1$, since in both NO and IO in the antineutrino flux component (giving rise to the IBD events, the most relevant channel) the factor $f_1$ containing $\tau_{\nu_1}$ is multiplied by the largest mixing matrix element, $|U_{e1}|^2$. 
The most constraining $95\%$~C.L. sensitivity we obtain is $b_{\nu_1}=\tau_{\nu_1}/m_\nu> 3.24\times 10^{6}$~s/eV, almost one order of magnitude better than the lower limits found from analyses of the SN1987A events. 
The sensitivity on $b_{\nu_2}$ is considerably improved from that obtained with solar neutrinos. The one on $b_{\nu_3}$ is also many orders of magnitude above that provided by atmospheric and long baseline  neutrino data, even if it is the worst sensitivity, due to the fact that in the antineutrino flux component (giving rise to the IBD events) the factor $f_3$ containing $\tau_{\nu_3}$ is multiplied by the smallest mixing matrix element, $|U_{e3}|^2$. Both $b_{\nu_2}$ and $b_{\nu_3}$ sensitivities are also better than the constraint obtained with high energy astrophysical neutrinos once all channels are combined.

\begin{figure*}[t]
    \centering
    \includegraphics[width=1.0\textwidth]{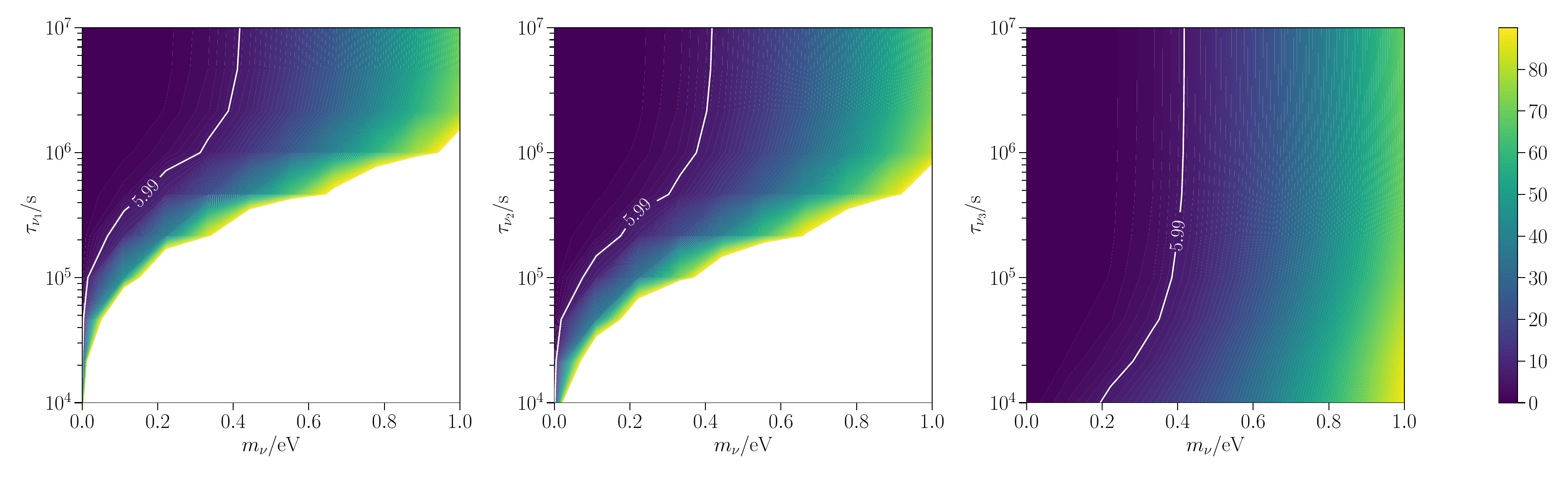}
    \caption{Contours at $95\%$ C.L. in the $m_\nu-\tau_\nu$ plane, obtained from the study of the combination of all the possible channels explored here, in the NO mass ordering scheme. Uncertainties on SN parameters are here neglected. Both effects of delay and decay, respectively, described by  Eq.~(\ref{eq:t_delay}) and Eq.~(\ref{eq:nu_decay}), are included in the analysis.}
    \label{fig:hk_comb_no}
\end{figure*}

\begin{figure*}[t]
    \centering
    \includegraphics[width=1.0\textwidth]{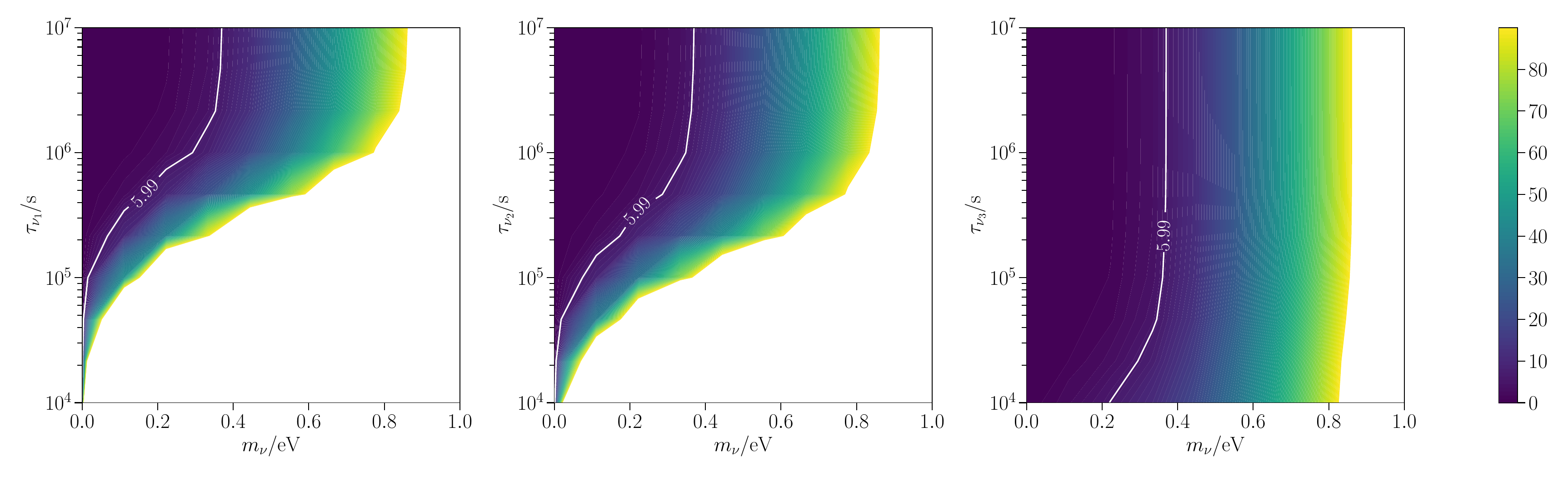}
    \caption{Same as Fig.~\ref{fig:hk_comb_no} but for the IO spectrum.}
    \label{fig:hk_comb_io}
\end{figure*}

\begin{figure*}
\begin{tabular}{c c c}
 \includegraphics[width = 0.33\textwidth]
{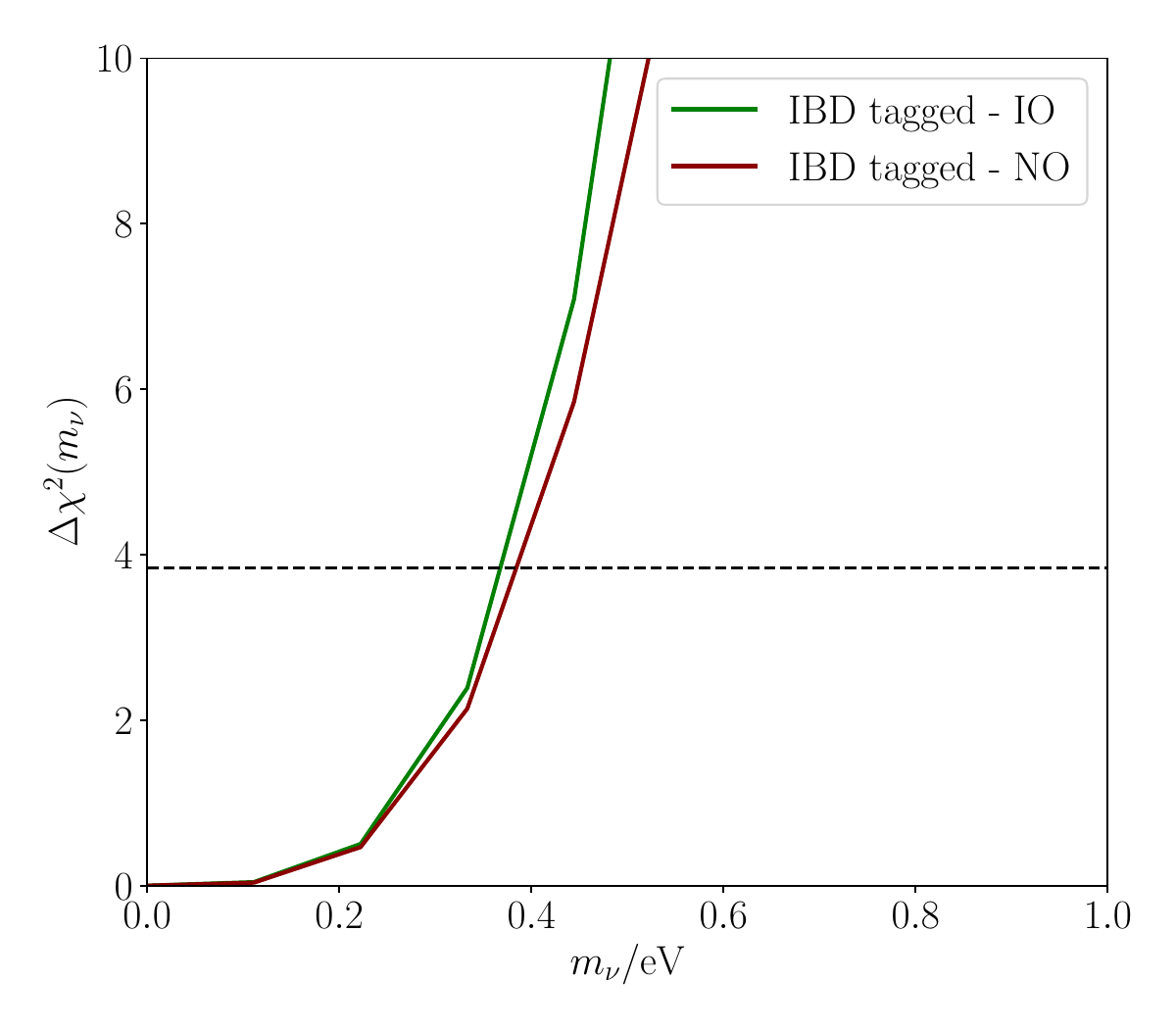}&
 \includegraphics[width = 0.33\textwidth]
{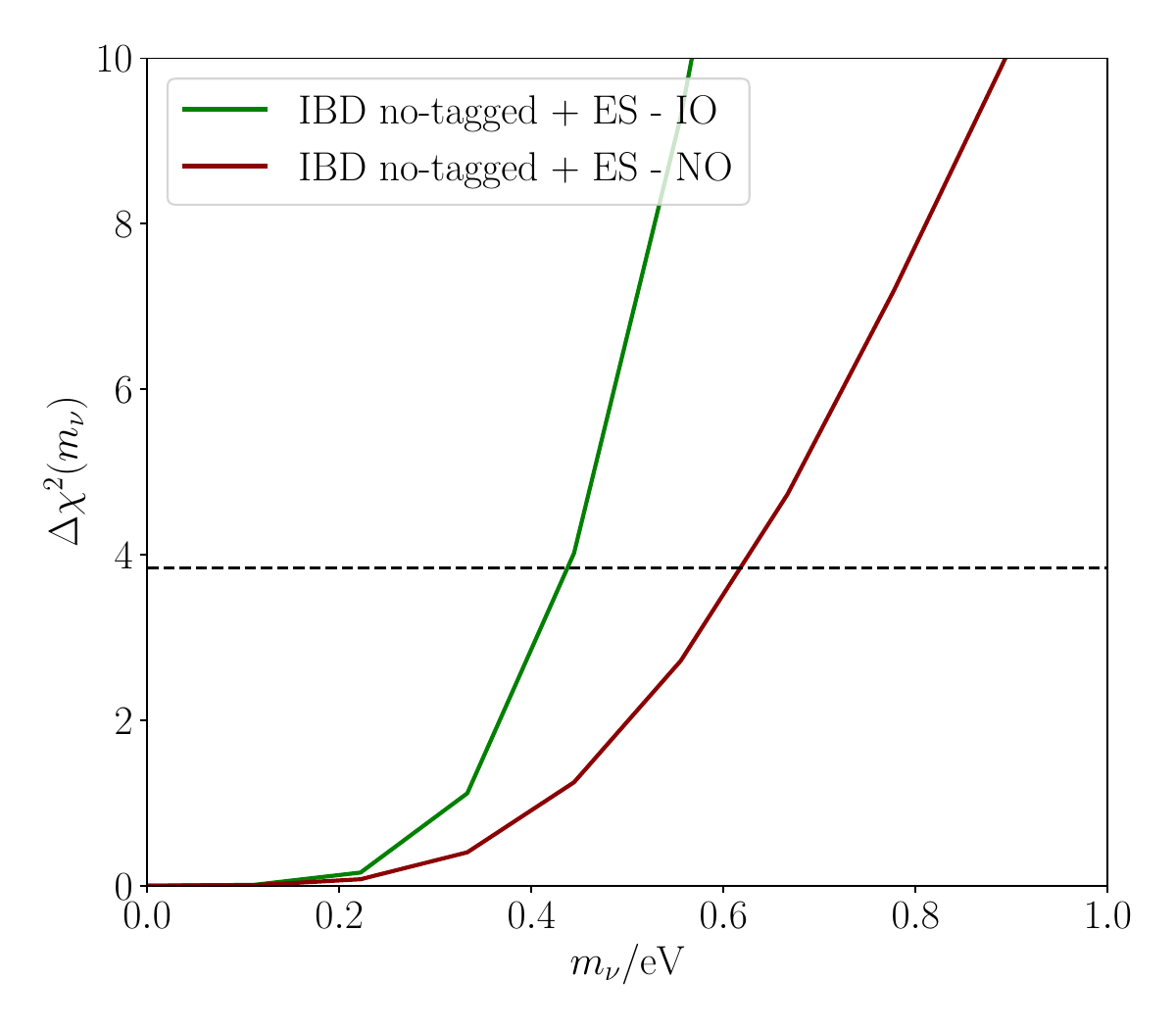}&
 \includegraphics[width = 0.33\textwidth]
{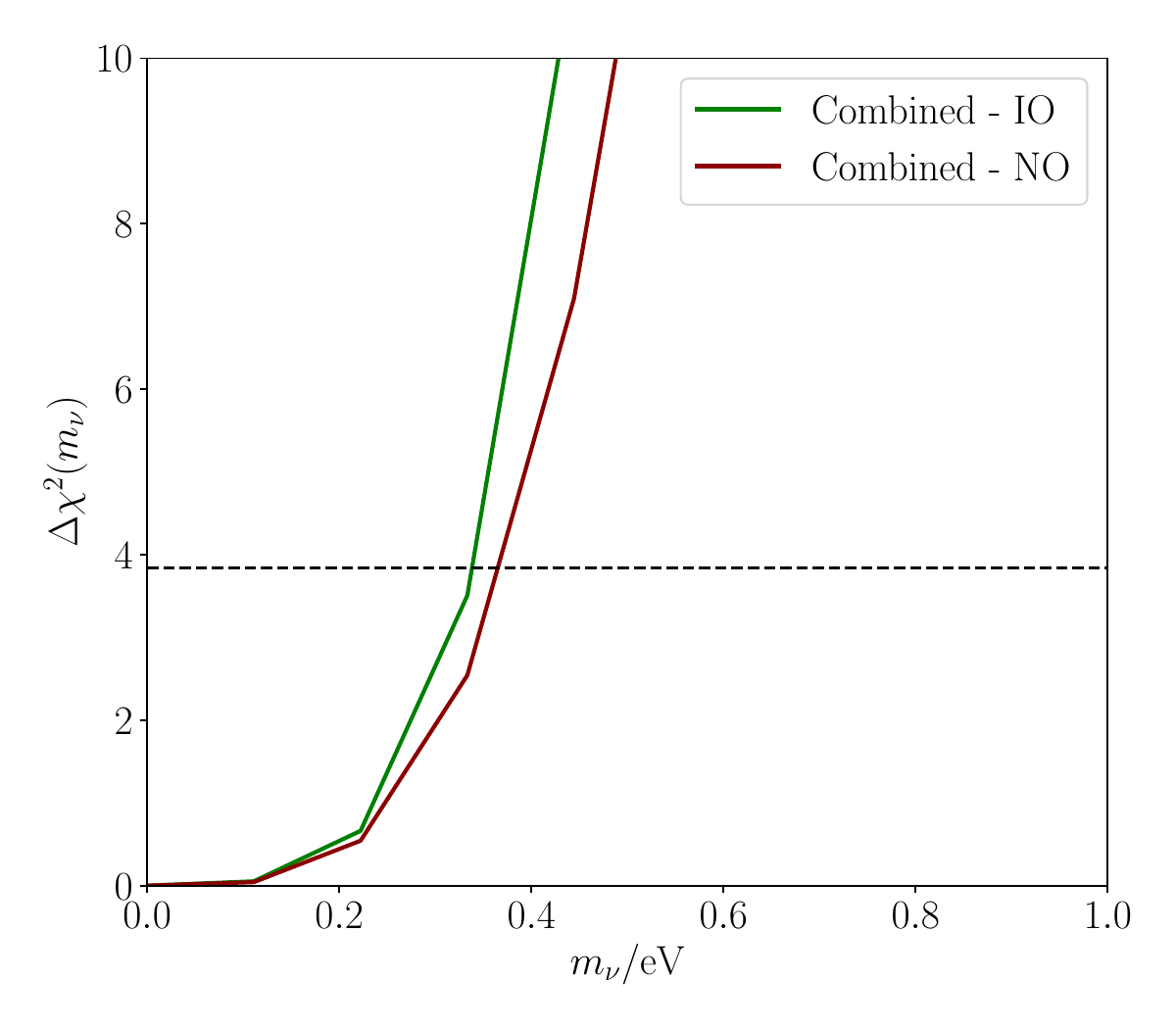}
\end{tabular}
\caption{From left to right panels: marginalized $\Delta \chi^2$ function versus the neutrino mass for the likelihood exploited in the time-delay signal in decaying neutrino scenarios from $90\%$ of the total expected IBD events, from $10\%$ of the total expected IBD events (not tagged by Gd) plus the contribution of all-flavor ES, and from both the tagged IBD and the ES events, see text for details. The horizontal lines depict the $95\%$~C.L. sensitivities.}
\label{fig:masstau}
\end{figure*}

Figures~\ref{fig:hk_comb_no} and \ref{fig:hk_comb_io} present the allowed contours in the ($m_{\nu_i}, \tau_{\nu_i}$) planes for NO and IO, respectively, arising from the combinations of all detection channels explored here (i.e. IBD tagged and untagged IBD + ES neutrino events). The white lines denote the $95\%$~C.L. allowed contours. 
Notice that the results barely change with the neutrino mass ordering. The marginalized limits and $\Delta \chi^2$ profiles for the mass are presented in Tab.~\ref{tab:mnu_bounds_both_eff} and Fig.~\ref{fig:masstau} respectively, for the different detection channels. 
The tightest bound on $m_\nu$ is obtained from the combination of all channels: $m_\nu<0.34~\ev$. It is very constraining and robust, as it is close to current cosmological neutrino mass limits, and it is independent of the precise neutrino mass ordering. This limit is also very close to the projected sensitivities from the beta-decay experiment KATRIN~\cite{Drexlin:2013lha}. 

\begin{figure*}
\begin{tabular}{c c c}
 \includegraphics[width = 0.33\textwidth]
{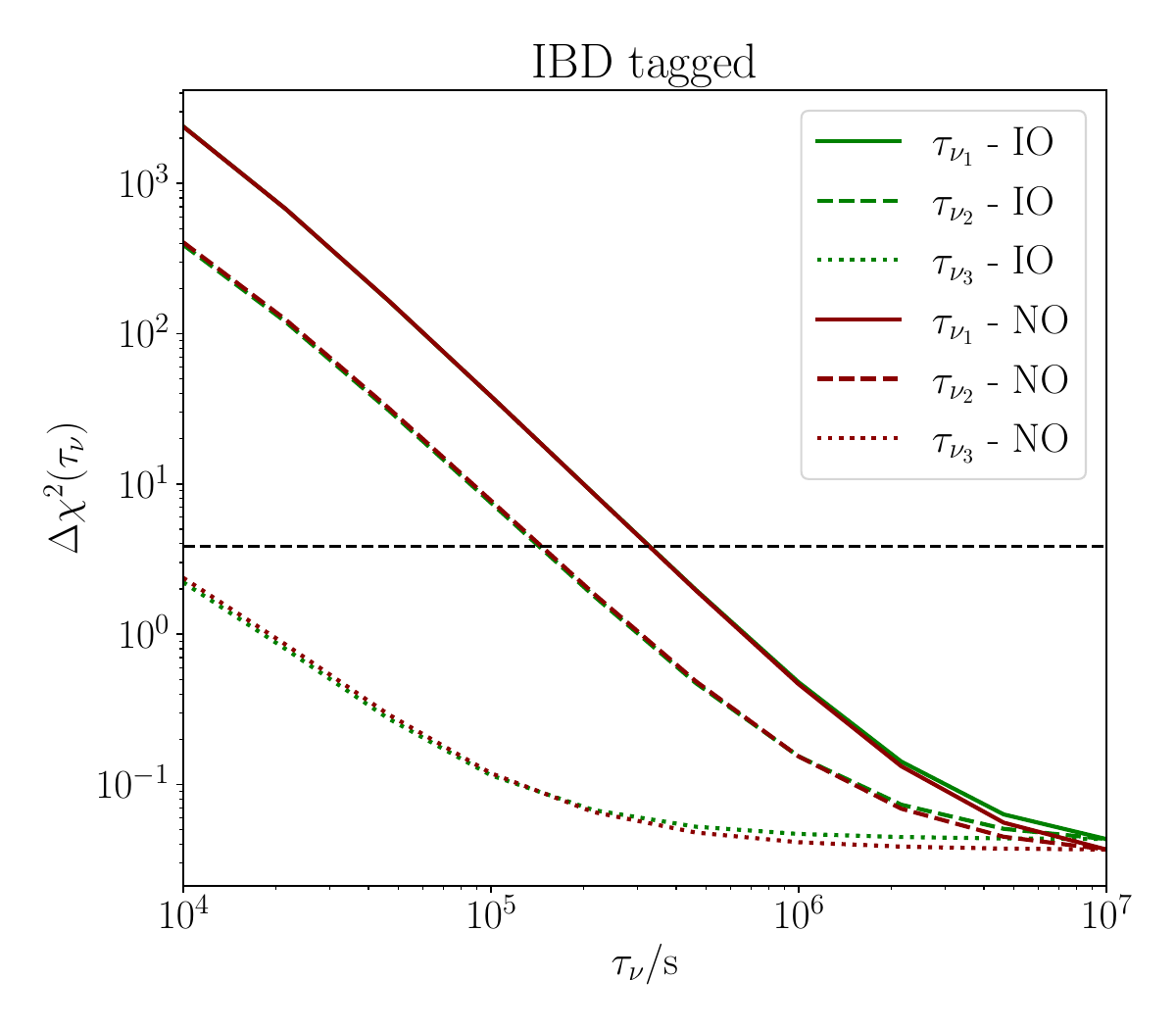}
 \includegraphics[width = 0.33\textwidth]
{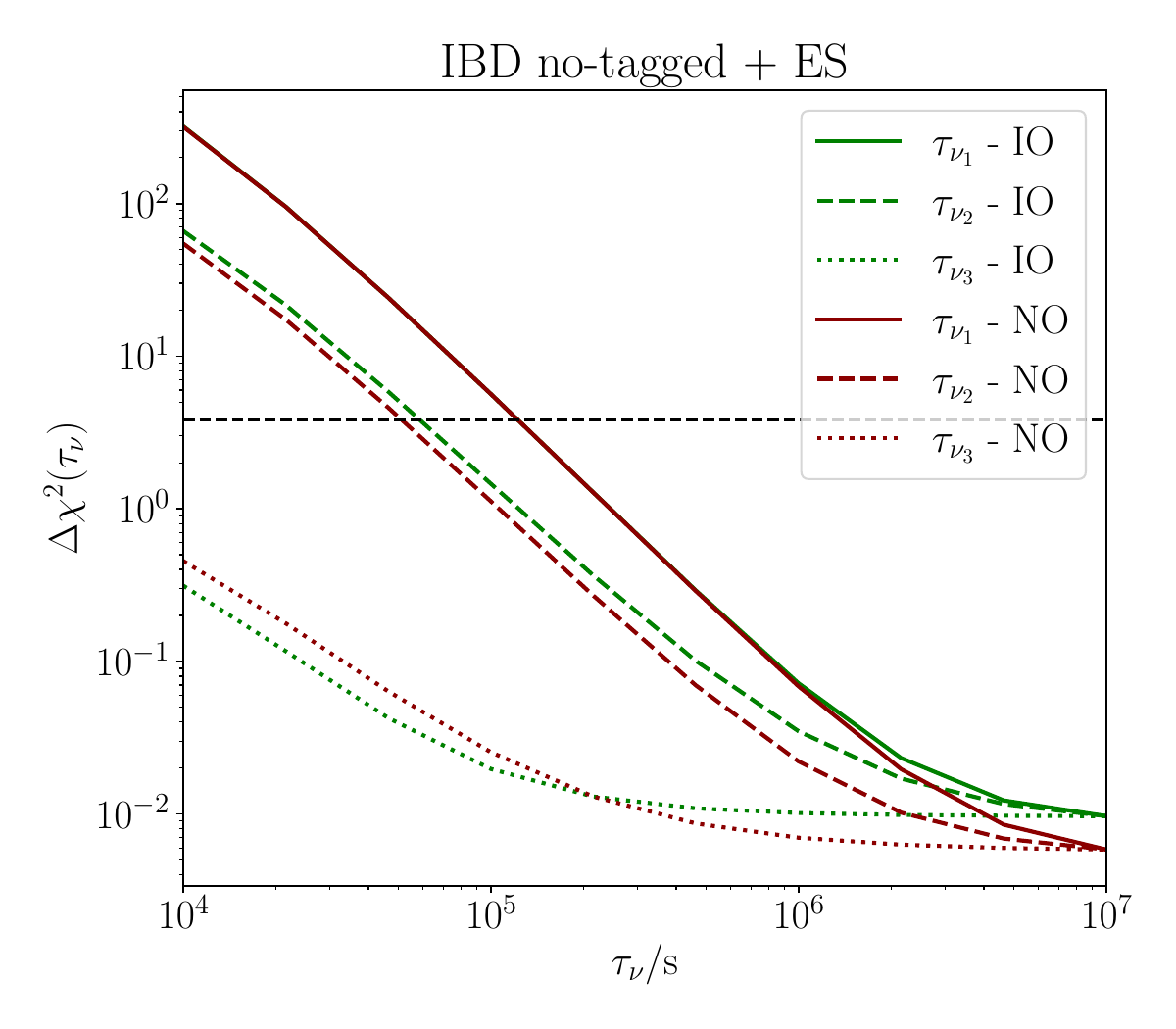}
 \includegraphics[width = 0.33\textwidth]
{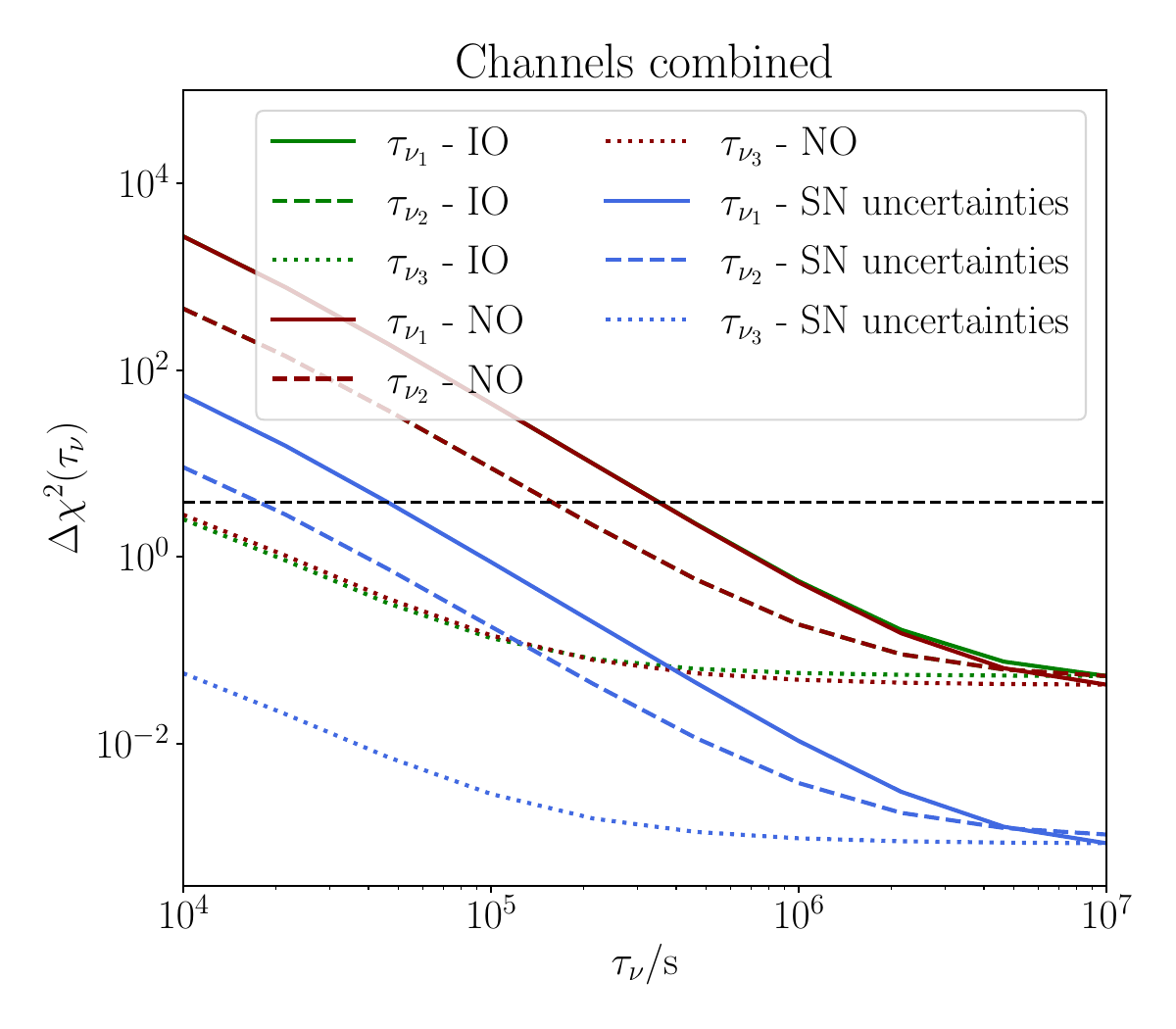}
\end{tabular}
\caption{From left to right panels: marginalized $\Delta \chi^2$ function versus the neutrino lifetime for the likelihood exploited in the time-delay signal in decaying neutrino scenarios from $90\%$ of the total expected IBD events, from $10\%$ of the total expected IBD events (not tagged by Gd) plus the contribution of all-flavor ES, and from both the tagged IBD and the ES events, see text for details. 
For the combined channels, NO case, we also illustrate the $\Delta \chi^2$ functions including Supernova neutrino flux uncertainties, see the main text for details. The horizontal lines depict the $95\%$~C.L. sensitivities.}
\label{fig:taumass}
\end{figure*}

Table~\ref{tab:tau_bounds_both_eff} and Fig.~\ref{fig:taumass} illustrate the marginalized limits and $\Delta \chi^2$ profiles, respectively, for the neutrino lifetimes $\tau_{\nu_i}$ of the different mass eigenstates. 
As in the case of the marginalized bounds for the neutrino mass, the $95\%$~C.L. constraints on $\tau_{\nu_i}$ do not depend on the neutrino mass ordering. Also in this case, as in the $b_{\nu_i}=\tau_{\nu_i}/m_{\nu_i}$ case shown in Tab.~\ref{tab:b_bounds_both_eff}, the bounds are very poor for the third mass eigenstate, especially in the case of untagged IBD plus ES events. This is expected, as the factor $f_3$ is always weighted by the smallest mixing matrix element $|U_{e3}|^2$. The tightest bound we find is $\tau_{\nu_1} > 4.13\times 10^5$~s at $95\%$~C.L. and after marginalizing over the neutrino mass.
To further assess the robustness of the limits derived here, we have also performed an additional analysis including the SN neutrino flux parameters as freely varying inputs. Namely, we have considered as extra parameters the mean energy and the pinching parameter in Eqs.~(\ref{eq:differential_flux}) and (\ref{eq:nu_energy_distribution})~\footnote{For the analysis concerning SN neutrino flux uncertainties, we have considered these extra two parameters to be flavor independent. In the absence of an underlying flavor-dependent parameterization,  this is a straightforward approach which allows us to quantify the overall size of the effect, discarding its particular flavor-behavior, which is not highly relevant for the aim of this study.}. 
The resulting upper bounds on the neutrino mass are mildly affected, as they do not strongly depend on the energy distribution, and are mainly extracted from the time-delay effect. 
In the case of the neutrino lifetime, since the constraints are mostly driven by the energy distribution profile, the $95\%$~C.L. lower bounds are modified, changing by one order of magnitude, as illustrated in the right panel of Fig.~\ref{fig:taumass}, where we show the marginalized $\Delta \chi^2$ functions also for the case in which SN flux uncertainties are included. 
As the results barely depend on the neutrino mass ordering, we only depict the NO case.
\begin{table*}[t]
  \centering
  \caption{$95\%$~C.L. lower bounds on $b_{\nu_i}=\frac{\tau_{\nu_i}}{m_{\nu_i}}$ ($\times 10^6~\mathrm{s/eV}$) for the different channels considered here, see the main text for details.} 
  \label{tab:b_bounds_both_eff}
  \begin{tabular}{c@{\hspace{10pt}}cccccc}
    \toprule
    \multirow{2}{*}{Channel} & \multicolumn{3}{c}{NO} & \multicolumn{3}{c}{IO} \\
    \cmidrule(lr){2-4} \cmidrule(lr){5-7}
    & $b_{\nu_1}$ & $b_{\nu_2}$ & $b_{\nu_3}$ & $b_{\nu_1}$ & $b_{\nu_2}$ & $b_{\nu_3}$ \\
    \midrule
    Gd-tagged IBD & $3.11$ & $1.31$ & $0.06$ & $3.10$ & $1.27$ & $0.06$ \\
    \midrule
    untagged IBD & \multirow{2}{*}{$1.17$} & \multirow{2}{*}{$0.47$} & \multirow{2}{*}{$<\!10^{-4}$} & \multirow{2}{*}{$1.17$} & \multirow{2}{*}{$0.55$} & \multirow{2}{*}{$<\!10^{-4}$} \\
    + total ES & & & & & & \\
    \midrule
    Combined & $3.24$ & $1.48$ & $0.07$ & $3.24$ & $1.48$ & $0.06$ \\
    \bottomrule
  \end{tabular}
\end{table*}
\begin{table}
    \centering
    \caption{$95\%$~C.L. sensitivity on $m_\nu$ (in $\ev$) upper bounds arising from the combined analysis of the effects due to neutrino delay and decay, for a SN at $D=10$~kpc.}
\label{tab:mnu_bounds_both_eff}
    \begin{tabular}{ccccc}
    \toprule
    Channel & & NO & & IO \\
    \midrule
    Gd-tagged IBD & & $0.38$ & & $0.37$ \\
    \midrule
    untagged IBD + total ES & & $0.62$ & & $0.44$ \\ 
    \midrule
    Combined & & $0.35$ & & $0.34$ \\
    \bottomrule 
    \end{tabular}
\end{table} 
\begin{table}[t]
  \centering
  \caption{$95\%$~C.L. sensitivity on $\tau_\nu$ ($\times 10^5~\mathrm{s}$) lower bounds resulting from the combined analysis of both the effects due to neutrino delay and decay, for a SN at $D=10$~kpc.} \begin{tabular}{c@{\hspace{12pt}}cccccc}
    \toprule
    \multirow{2}{*}{Channel} & \multicolumn{3}{c}{NO} & \multicolumn{3}{c}{IO} \\
    \cmidrule(lr){2-4} \cmidrule(lr){5-7}
    & $\tau_{\nu_1}$ & $\tau_{\nu_2}$ & $\tau_{\nu_3}$ & $\tau_{\nu_1}$ & $\tau_{\nu_2}$ & $\tau_{\nu_3}$ \\
    \midrule
    Gd-tagged IBD & $3.94$ & $1.77$ & $0.07$ & $3.94$ & $1.74$ & $0.07$ \\
    \midrule
    untagged IBD  & \multirow{2}{*}{$1.48$} & \multirow{2}{*}{$0.58$} & \multirow{2}{*}{$<\!10^{-3}$} & \multirow{2}{*}{$1.48$} & \multirow{2}{*}{$0.71$} & \multirow{2}{*}{$<\!10^{-3}$} \\
    + total ES & & & & & & \\
    \midrule
    Combined & $4.12$ & $1.86$ & $0.08$ & $4.13$ & $1.87$ & $0.08$ \\
    \bottomrule
  \end{tabular}
  \label{tab:tau_bounds_both_eff}
\end{table}

%
%
%

\section{Conclusions}
\label{sec:conclusions}
Next generation neutrino detectors will be very sensitive core-collapse SN neutrino observatories. A huge number of events is expected in the Hyper-Kamiokande (HK) water Cherenkov detector, provided the SN explosion occurs in our galaxy within a few tens of kiloparsecs. With thousands of neutrino events expected, we have explored here the potential of HK to 
measure the neutrino mass and lifetime. Assuming that the SN is located at $10$~kpc from us, we have simulated the neutrino events from two different channels, namely, IBD electron anti-neutrino tagged events due to the presence of Gadolinium, which represent $90\%$  of the total IBD events, and untagged IBD events plus those from all flavors of neutrinos ad antineutrinos coming from ES processes with electrons in the detector. 
 Despite the lower statistics coming from the ES channel compared to the tagged-IBD one, its contribution is of great importance in order to set constraints on the absolute value of neutrino mass via time delays, as one would be sensitive to the neutronization peak present in the SN electron neutrino flux. 
 The gain when exploiting that channel is twofold. On the one hand, the sensitivity limits are improved by a factor of two. Secondly, the bounds are almost independent of the neutrino mass ordering. All in all, we find a $95\%$~C.L. sensitivity limit of $\sim$ 0.4-0.5~eV, very close to future sensitivities from laboratory neutrino mass searches. 
We have also explored the effect of neutrino decays. It is possible therefore not only to constrain $\tau_\nu/m_\nu$ but also $\tau_\nu$ and $m_\nu$ independently. From what concerns the typical decay parameter $\tau_\nu/m_\nu$, we find sensitivities that are one order of magnitude better than the limits reported from SN1987A. 
When extracting simultaneously $m_\nu$ and $\tau$, we obtain $m_\nu\lesssim 0.3$~eV and $\tau_\nu>4\times 10^5$~s, both at $95\%$~C.L. 
Finally, we have reassessed the robustness of our limits by including the possible uncertainties in the SN neutrino fluxes, finding that in the case of the neutrino mass, the lower limits remain stable while the limits on $\tau$ are worse.  
The current study therefore clearly states the rich potential of future neutrino detectors in constraining neutrino properties robustly by exploiting core-collapse SN neutrino events.

\begin{acknowledgments}
The authors would like to thank F.~Capozzi and M.~Sorel for very useful improvements and suggestions.
This work has been supported by the Spanish  grant PID2020-113644GB-I00 and by the European Union’s Framework Program for Research and Innovation Horizon 2020 (2014–2020) under grant H2020-MSCA-ITN-2019/860881-HIDDeN and SE project ASYMMETRY (HORIZON-MSCA-2021-SE-01/101086085-ASYMMETRY) and well as by the Generalitat Valenciana grants PROMETEO/2019/083 and CIPROM/2022/69.
\end{acknowledgments}

\bibliography{biblio}
\end{document}